# Attention-Based Deep Neural Networks for Detection of Cancerous and Precancerous Esophagus Tissue on Histopathological Slides

Naofumi Tomita, MS[1]; Behnaz Abdollahi, PhD[1]; Jason Wei, BS[1,2]; Bing Ren, MD[3]; Arief Suriawinata, MD[3], Saeed Hassanpour, PhD[1,2,4] *

[1]Department of Biomedical Data Science, Geisel School of Medicine at Dartmouth, Hanover, NH 03755, USA

[2]Department of Computer Science, Dartmouth College, Hanover, NH 03755, USA

[3]Department of Pathology and Laboratory Medicine, Dartmouth-Hitchcock Medical Center, Lebanon, NH 03756, USA

[4]Department of Epidemiology, Geisel School of Medicine at Dartmouth, Hanover, NH 03755, USA

# ABSTRACT

**Importance:** Deep learning–based methods, such as the sliding window approach for cropped-image classification and heuristic aggregation for whole-slide inference, for analyzing histological patterns in high-resolution microscopy images have shown promising results. These approaches, however, require a laborious annotation process and are fragmented.

**Objective:** To evaluate a novel deep learning method that uses tissue-level annotations for high-resolution histological image analysis for Barrett esophagus (BE) and esophageal adenocarcinoma detection.

**Design, Setting, and Participants:** This diagnostic study collected deidentified high-resolution histological images (N = 379) for training a new model composed of a convolutional neural network and a grid-based attention network. Histological images of patients who underwent endoscopic esophagus and gastroesophageal junction mucosal biopsy between January 1, 2016, and December 31, 2018, at Dartmouth-Hitchcock Medical Center (Lebanon, New Hampshire) were collected.

**Main Outcomes and Measures:** The model was evaluated on an independent testing set of 123 histological images with 4 classes: normal, BE-no-dysplasia, BE-with-dysplasia, and adenocarcinoma. Performance of this model was measured and compared with that of the current state-of-the-art sliding window approach using the following standard machine learning metrics: accuracy, recall, precision, and F1 score.

**Results**: Of the independent testing set of 123 histological images, 30 (24.4%) were in the BE-no-dysplasia class, 14 (11.4%) in the BE-with-dysplasia class, 21 (17.1%) in the adenocarcinoma class, and 58 (47.2%) in the normal class. Classification accuracies of the proposed model were 0.85 (95% CI, 0.81-0.90) for the BE-no-dysplasia class, 0.89 (95% CI, 0.84-0.92) for the BE-with-dysplasia class, and 0.88 (95% CI, 0.84-0.92) for the adenocarcinoma class. The proposed model achieved a mean accuracy of 0.83 (95% CI, 0.80-0.86) and marginally outperformed the sliding window approach on the same testing set. The F1 scores of the attention-based model were at least 8% higher for each class compared with the sliding window approach: 0.68 (95% CI, 0.61-0.75) vs 0.61 (95% CI, 0.53-0.68) for the normal class, 0.72 (95% CI, 0.63-0.80) vs 0.58 (95% CI, 0.45-0.69) for the BE-no-dysplasia class, 0.30 (95% CI, 0.11-0.48) vs 0.22 (95% CI, 0.11-0.33) for the BE-with-dysplasia class, and 0.67 (95% CI, 0.54-0.77) vs 0.58 (95% CI, 0.44-0.70) for the adenocarcinoma class. However, this outperformance was not statistically significant.

**Conclusions and Relevance:** Results of this study suggest that the proposed attention-based deep neural network framework for BE and esophageal adenocarcinoma detection is important because it is based solely on tissue-level annotations, unlike existing methods that are based on regions of interest. This new model is expected to open avenues for applying deep learning to digital pathology.

## BACKGROUND

### Esophageal Cancer

Barrett's esophagus (BE) is a transformation of the normal squamous epithelium of the esophagus into metaplastic columnar epithelium.[1] BE is important because it predisposes patients to the increased risk of adenocarcinoma of the esophagus and gastroesophageal junction.[2,3] Compared to the general population, patients with BE have a 30 to 125 times higher risk of cancer.[4] The average 5-year survival rate for esophageal adenocarcinoma (EAC) in the U.S. is less than 15%.[5] Furthermore, the incidence of EAC increased dramatically in the U.S. over three decades.[6-10] Histologic diagnosis of BE requires the identification of metaplastic columnar epithelium with goblet cells (i.e., intestinal metaplasia).[11] Evaluating the development of the premalignancy and malignancy in BE shows a moderate interobserver variability, with an average kappa coefficient of less than 0.50 even among subspecialized gastrointestinal pathologists.[12]

### Deep Learning for Pathology Image Analysis

In the field of digital pathology, tissue slides are scanned as high-resolution images, which can have sizes up to 10,000×10,000 pixels. This high resolution is necessary because each slide contains thousands of cells, for which the cellular structures must be visible in order to identify regions of the tissue with diseases or lesions. However, the size of lesions is often relatively small, typically around 100×100 pixels, as most of the tissue areas in a given slide are normal. Therefore, the decisive regions of interest (ROIs) containing lesions usually comprise much less than one percent of the entire scanned tissue area. Even for highly trained pathologists, localizing these lesions for the classification of the whole slide is time consuming and often error-prone.

In recent years, deep learning has made considerable advances in classifying microscopy images. The most common approach in this domain involves a sliding window aproach for cropped-image classification, followed by statistical methods of aggregation for whole-slide inference.[13-23] In this approach, pathologists annotate bounding boxes (i.e., ROI) on whole slides in order to train a classifier on small cropped-images, typically of sizes in the range of 200×200 pixels to 500×500 pixels. For evaluating a whole slide, this cropped-image classifier is applied to extracted windows from the image, and then a heuristic, often developed in conjunction with a domain-expert pathologist, is used to determine how the distribution of cropped-image classification scores translates into a whole-slide diagnosis.

However, there are many limitations to this sliding window approach. The first is that since cropped-image classifiers are needed, all images in the training set must be annotated by pathologists with bounding boxes around each ROI. In addition, developing a heuristic for aggregating cropped-image classifications, which requires pathologists' insight, is dependent on the nature of the classification task and is not widely scalable. Finally, in the sliding window approach, cropped-images are classified independently of their neighbors and whole-slide classification does not consider correlations between neighboring windows. In order to overcome these limitations in this work, we propose to use an attention mechanism where the ROI is mined from high-resolution slides without explicit supervision.

Our work is inspired by attention models applied to regular image analysis tasks, especially image captioning.[24,25] Attention mechanisms are described as a part of the prediction module that sequentially selects subsets of input to be processed.[24] Although this definition is not applicable to non-sequential tasks, the essence of attention mechanisms can be restructred for neural networks to generate a dynamic representation of features through by weighting them to capture a holistic context of input. Unlike hard attention, where an ROI is selected by a stochastic sampling process, soft attention generates a non-discrete attention map that pays fractional attention to each region and produces better gradient flow, and thus is easier to optimize. Recent advancement of soft attention enabled end-to-end training on convolutional neural network (CNN) models.[26-29] For example, spatial transformer networks capture high-level information from inputs to derive affine transformation parameters, which are subsequently applied to spatial invariant input for a CNN.[29] For semantic segmentation tasks, the attention mechanism is applied to learn multi-scale features.[26] Residual attention networks use soft attention masks to extract features in different granularities.[28] To analyze images in detail, a top-down recurrent attention CNN has been proposed.[27] To put our work into perspective, our proposed model is based on the soft attention mechanism in feature space, but is designed for the classification of high-resolution images that are not typically encountered in the field of computer vision. There have been several applications of the attention mechanism in the medical domain, such as using soft attention to generate masks around lesion areas on CT images[30] and employing recurrent attention models fused with reinforcement learning to locate lung nodules[31] or enlarged hearts[32] in chest radiography images. In pathology, recorded navigation of pathologists has been used as attention maps to detect carcinoma.[33] A soft attention approach has been deployed in two parallel networks for the classification of thorax disease.[30] Although we draw inspiration from this work, our approach differs in that it provides a novel framework to directly reuse extracted features in a single attention network.

In this paper, we present a model that uses a convolutional attention-based mechanism to classify microscopy images. Our approach has three major advantages over the existing method. First, our model dynamically identifies ROIs in a high-resolution image and makes a whole-slide classification based on analyzing only these selected regions. This is analogous to how pathologists examine slides under the microscope. Second, the proposed model is trainable end-to-end with only tissue-level labels. All components of our model are optimized through backpropagation. Unlike the current sliding window approach, our model does not need bounding box annotations for ROIs or a pathologist's insight for heuristic development.

Lastly, the model architecture is flexible with regard to input size for images. Inspired by fully convolutional network philosophy,[34] our grid-based attention module uses a 3D convolution operation that does not require a fixed size input grid. The input size can be any rectangular shape that fits in the memory of graphic processing units (GPUs), which all modern deep learning frameworks utilize to accelerate computations.

**MATERIALS AND METHODS**

**Dataset**

For this study, whole-slide images were collected from patients who underwent endoscopic esophagus and gastroesophageal junction mucosal biopsy since 2017 at Dartmouth-Hitchcock Medical Center (DHMC), a tertiary academic medical center in Lebanon, New Hampshire. The use of data collected in this project is approved by the Dartmouth Institutional Review Board (IRB) and the research conducted in this paper is in compliance with the World Medical Association Declaration of Helsinki on Ethical Principles for Medical Research Involving Human Subjects. A Leica Aperio scanner was used to digitize H&E-stained whole-slide images at 20× magnification. We had a total of 180 whole-slide images, of which 116 (i.e., 64% of the dataset) were used as the development set and 64 (i.e., 36% of the dataset) were used as the test set. 20% of the development set whole-slide images were reserved for validation. Of note, these whole-slide images can cover multiple pieces of tissue. Therefore, the whole-slide images were separated into 256 high-resolution images later in our preprocessing step, each image covering a single piece of tissue.

In order to determine labels for whole-slide images and to train the existing state-of-the-art sliding method as our baseline, bounding boxes around lesions in these images were annotated by two expert pathologists from the Department of Pathology and Laboratory Medicine at DHMC. We considered these labels as the reference standard, as any disagreements in annotation were resolved through further discussion among annotators and consultation with a senior domain-expert pathologist. These bounding boxes were not needed in training our proposed attention-based model.

This project used categories of esophageal cancer as defined by the Vienna classification system.[35] The classification of normal includes normal squamous epithelium, normal squamous and columnar junctional epithelium, and normal columnar epithelium. BE-no-dysplasia includes Barrett's esophagus negative for dysplasia. Barrett's esophagus is defined by columnar epithelium with goblet cells (intestinal metaplasia) and preservation of orderly glandular architecture of the columnar epithelium with surface maturation. BE-with-dysplasia includes low-grade dysplasia (noninvasive low-grade neoplasia) and high-grade dysplasia (noninvasive high-grade neoplasia). Columnar epithelium with low-grade dysplasia is characterized by nuclear pseudostratification, mild to moderate nuclear hyperchromasia and irregularity, and the cytologic atypia extending to the surface epithelium. High-grade dysplasia demonstrates marked cytologic atypia including loss of polarity, severe nuclear enlargement and hyperchromasia, numerous mitotic figures, and architectural abnormalities such as lateral budding, branching, villous formation, as well as variation of the size and shape of crypts. Adenocarcinoma includes invasive carcinoma (intramucosal carcinoma and submucosal carcinoma and beyond) and high grade dysplasia suspicious for invasive carcinoma. Cases in the adenocrcinoma category may present the following features: single cell infiltration, sharply angulated glands, small glands in a back-to-back pattern, confluent glands, cribriform/solid growth, ulceration occurring within high-grade dysplasia, dilated dysplastic glands with necrotic debris, or dysplastic tubules incorporated into overlying squamous epithelium.

**Methodology**

Our proposed approach has two steps, which is shown in Figure 1. The first step is grid-based feature extraction from the high-resolution image, where we analyze each grid cell in the whole slide to generate a feature map (Figure 1.a-b). In the second step, we apply our proposed attention mechanism on the extracted features for slide classification (Figure 1.c). Notably, the feature extractor is jointly optimized across all the grid cells along with the attention module in an end-to-end fashion. In the end-to-end training pipeline, the cross-entropy loss over all classes is computed on classification predictions. The loss is backpropagated to optimize all parameters in the network without any manual adjustment for attention modules. Of note, our model does not need bounding box annotations around ROIs, and all optimization is done with respect to only the labels at the tissue-level. Further details of the model architecture of the grid-based feature extraction and attention-based classification are provided in the Supplement Material section.

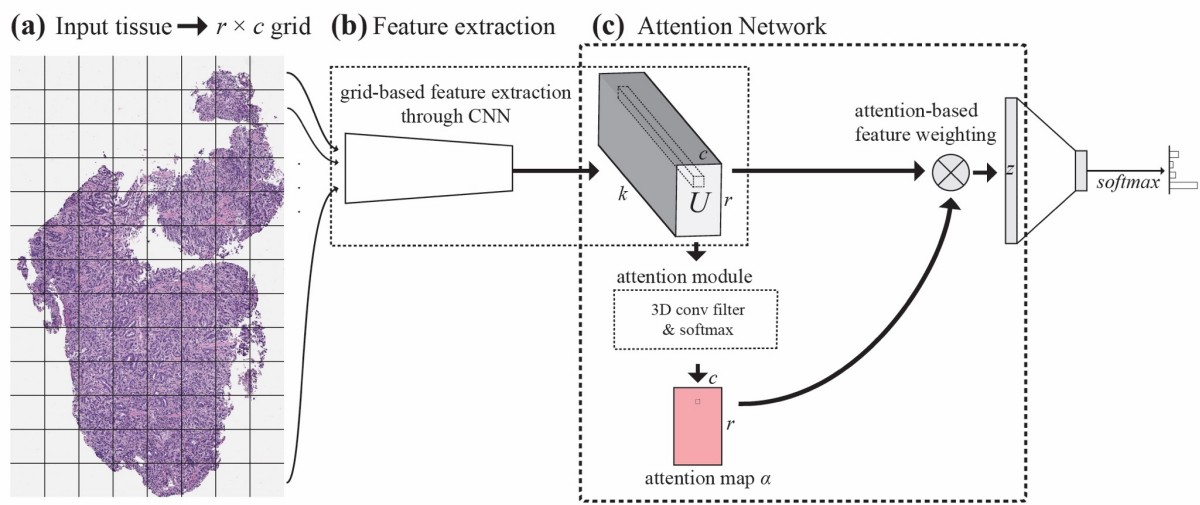

**Figure** *1* <Please see the figure legend at the end of the manuscript.>

**EXPERIMENTS**

To evaluate our attention-based classification model for high-resolution microscopy images, we applied our method to high-resolution scanned slides of tissues endoscopically removed from patients at risk of esophageal cancer. We compared the performance results of our proposed model to those generated by the state-of-the-art sliding window method.[22]

For preprocessing, we removed the white background from the slides and extracted only regions of the images that contained tissue. eFigure 1a shows a typical whole-slide image from our dataset. These whole-slide images can cover multiple pieces of tissue, so we separated them into large images with an average size of 5,131×5,875 pixels, with each only covering a single piece of tissue. Every image was given an overall label based on the labels of its lesions. If multiple lesions with different classes were present, we used the class with the highest risk as the corresponding label, as that lesion would have the highest impact clinically. If no abnormal lesions were found in an image, it was assigned to the normal class. After this preprocessing step, each image was assigned to one of our four classes: normal, BE-no-dysplasia, BE-with-dysplasia, and adenocarcinoma (eFigure 1b). Our dataset included 379 images after preprocessing. One third of the dataset was reserved for testing. To avoid possible data leakage, tissues extracted from one whole-slide image were all placed into the same set of images when the development and test sets were split. Table 1 summarizes our test set.

**Sliding Window Baseline**

In order to compare our model to previous methods for high-resolution image analysis, we implemented the current state-of-the-art sliding window method.[22] In this method, we used our annotated bounding box labels to generate small cropped-images of size 224×224 pixels for training a cropped-image classifier. For preprocessing, we normalized the color channels and performed standard data augmentation, including color jittering, random flips, and rotations. For training, we initialized ResNet-18 with the MSRA initialization.[36] We optimized the model with a cross-entropy loss function for 100 epochs, employing standard weight regularization techniques and learning rate decay. We trained our cropped-image classifier to predict the class of any given window on a high-resolution image. For whole-slide inference, we performed a grid search over our validation set to find optimal thresholds for filtering noise. Then, we consulted separately with two pathologists to develop heuristics for aggregating cropped-image predictions. We chose the thresholds and heuristic that performed the best on the validation set and applied that to the whole-slide images in the test set.

**Attention Model**

We implemented our attention model as described in the Methodology section. Given the size of features extracted from the ResNet-18 model, we used 512×3×3 3D convolutional filters in the attention module, with implicit zero-padding of (0, 1, 1) for depth, height, and width dimensions. We employed 64 of these filters to increase the robustness of the attention module, as patterns in the feature space are likely too complex to be recognized and attended by a single filter. To avoid overfitting and encourage each filter to capture different patterns, we regularized the attention module by applying dropout[37] with $p = 0.5$ after concatenating all the feature vectors. We initialized the entire network with the MSRA initialization for convolutional filters,[36] unit weight and zero-bias for batch normalizations,[38] and the Glorot initialization for fully connected layers.[39] Notably, only the cross-entropy loss against class labels was used in training. Other information such as the location of bounding boxes was not given to the network as guidance to optimal attention maps. Our model identified such ROIs automatically.

We first initialized our feature extraction network with weights pretrained on the ImageNet dataset.[40] Input for the network were extracted grid cells of 492×492 pixels that were resized to 224×224 pixels. We normalized the input values by the mean and standard deviation of pixel values computed over all tissues in the traning set. In our training, the last fully connected layer of the network was removed, and all residual blocks except for the last one were frozen, serving as a regularization mechanism.

We trained the entire network on large, high-resolution images. For data augmentation, we applied random rotation and random scaling with a scaling factor between 0.8 and 1.2 during training. We used the Adam optimizer with an initial learning rate of 1e-3, decaying by 0.95 after each epoch, and reset the learning rate to 1e-4 every 50 epochs in a total of 200 epochs, similar to the cyclical learning rate.[41,42] We set the mini batch size to 2 to maximize the utilization of memory on our Nvidia Titan Xp GPU. The model was implemented in PyTorch.[43] At testing, the network took 0.34s on average for the analysis of a high-resolution image.

**EVALUATION AND RESULTS**

We evaluated trained models on the test set and analyzed the classification performance from both quantitative and qualitative aspects. As a baseline, we referred to results from using the sliding-window method[22] for this classification task, which was trained on the same data split, but with annotated bounding boxes. For quantitative evaluation, we used four standard metrics for classification: accuracy, recall, precision, and F1 score using a one-vs-rest strategy. Our classification results on the test set are summarized in Table 2. Compared to the baseline, our model achieved better accuracy and F1 score in all classes. Especially for F1 score, which is the harmonic mean of precision and recall, our model outperformed the baseline approach by at least 8% for each class; however, this outperformance was not significant at the 0.05 level of statistical significance. Quantitative analysis showed an exemplary performance of our model on the normal, BE-no-dysplasia, and adenocarcinoma classes. However, both our attention model and the baseline model did not perform as well in identifying images of the class BE-with-dysplasia. As shown in the confusion matrix in Figure 2, most samples of BE-with-dysplasia images that were misclassified by the attention model are predicted as normal tissue. This is likely because BE-with-dysplasia is the least frequent class in our dataset, comprising only 11% of images. For further comparison, the Receiver Operating Characteristic (ROC) curves of both models for each class are plotted in Figure 3. The attention model is trained without ROI annotations, yet achieved compelling area under the ROC curve (AUC) values for each class.

| Ground Truth | | Sliding Window[22] | Attention Model |
|---|---|---|---|
| Normal | Accuracy | 0.63 | 0.70 |
| | Recall | 0.62 | 0.69 |
| | Precision | 0.60 | 0.68 |
| | F1 Score | 0.61 | 0.68 |
| BE-no-dysplasia | Accuracy | 0.85 | 0.85 |
| | Recall | 0.43 | 0.77 |
| | Precision | 0.87 | 0.68 |
| | F1 Score | 0.58 | 0.72 |
| BE-with-dysplasia | Accuracy | 0.72 | 0.89 |
| | Recall | 0.36 | 0.21 |
| | Precision | 0.16 | 0.50 |
| | F1 Score | 0.22 | 0.30 |
| Adenocarcinoma | Accuracy | 0.87 | 0.88 |
| | Recall | 0.52 | 0.71 |
| | Precision | 0.65 | 0.63 |
| | F1 Score | 0.58 | 0.67 |
| Mean | Accuracy | 0.76 | 0.83 |
| | Recall | 0.48 | 0.60 |
| | Precision | 0.57 | 0.62 |
| | F1 Score | 0.50 | 0.59 |

**Table 1.** Classification results for our test set. We assessed the model's performance in terms of accuracy, recall, precision, and F1 score. Results are rounded to two decimal places. Our method outperforms the sliding window baseline in both accuracy and F1 score for all classes.

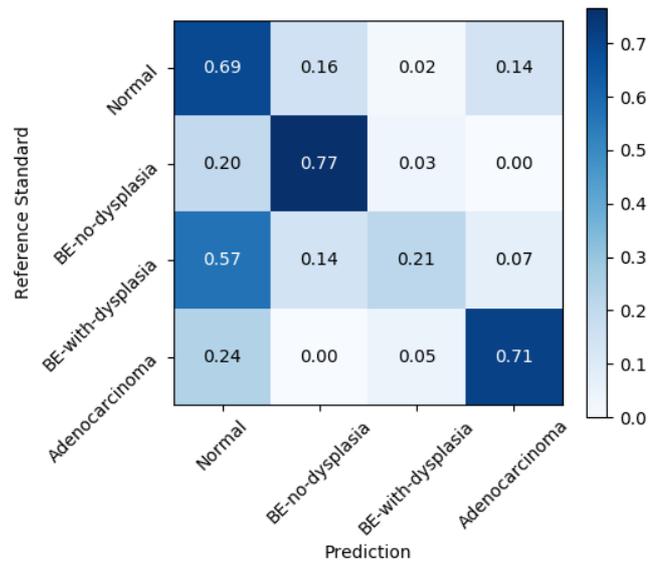

**Figure** *2* <Please see the figure legend at the end of the manuscript.>

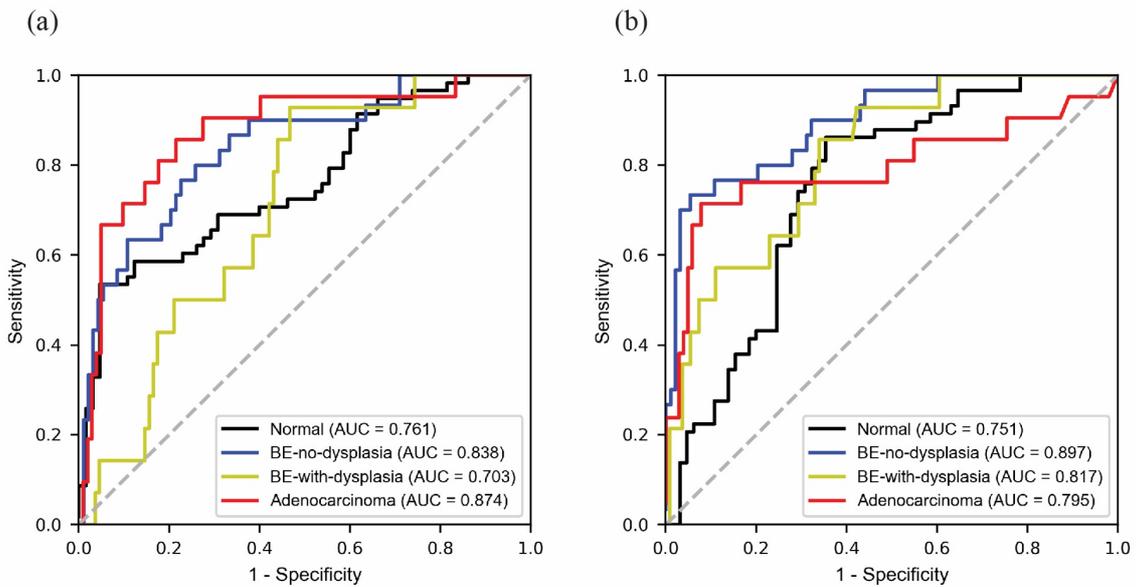

**Figure 3** *<Please see the figure legend at the end of the manuscript.>*

The attention maps generated for all the testing images were visualized to verify the attention mechanism in our model. We present characteristic examples for the adenocarcinoma class in Figure 4. The distributions of the attention weights highlighted across different classes indicate that the attention module looks for specific features in the adenocarcinoma class (Figure 4.d). For images without the target features, the attention weights are low over all regions (Figure 4.a-b). In Figure 4.c, we observe that the attention map is clinically on-target and is focused on specific regions in which BE-with-dysplasia progresses to adenocarcinoma as neoplastic epithelia begin to invade the muscularis mucosae.[44]

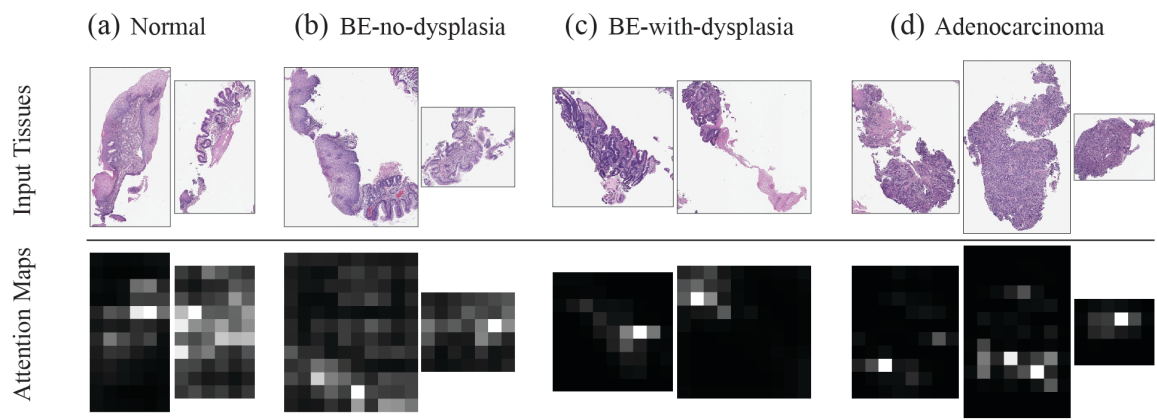

**Figure 4** <Please see the figure legend at the end of the manuscript.>

**DISCUSSION**

Our results demonstrated the detection of BE/EAC using an attention-based deep learning architecture. The classification performance on our dataset is higher than that of the state-of-the-art sliding window model. This is significant because our proposed model only needs reference labels per tissue, while the existing sliding window models require bounding box annotations for each ROI in a tissue. While both models use a ResNet18 model for feature extraction, the attention mechanism of our model further directs the information flow and forces the network to identify local features that are useful for classification. Our proposed architecture is directly applicable to high-resolution images without resizing due to its flexible input design. Considering the time and resources required for annotating microscopy images, fewer requirements for these annotations would facilitate image analysis research and development on these images. Particularly, tissue-level annotations required for training our architecture can potentially be retrieved through searches in pathology reports associated with microscopy images. The proposed classification scheme is potentially applicable to histology images of other diseases for which training data is scarce or bounding box annotations are not available. Of note, our model is also the first to automate detection of BE and EAC on histopathology slides using a deep learning model.

Our proposed approach has some limitations. In terms of our dataset, one limitation is that all experiments were conducted on slides collected from a single medical center and scanned with the same equipment. Another is that our dataset is still relatively small in comparison to conventional datasets in deep learning; in particular, the number of slides of BE-with-dysplasia was small, resulting in lower performance for that class. In order to evaluate the robustness and generalizability of our novel approach, further verification with different classification tasks and larger datasets from various institutions is required and will be pursued in future work.

Furthermore, even with our method, which is built to analyze entire tissue regions, current GPUs do not have enough memory capacity to process very large images. For such slides, we can divide the tissue area into manageable sub-tissue images. Alternatively, the feature extractor, which is the largest source of memory consumption in our approach, can be optimized to address this issue. The ResNet-18 architecture used in our model achieved high performance with a relatively low number of parameters. However, there is still room for further reduction of parameters while maintaining high performance, which we will pursue in future work.

# CONCLUSION

We presented a new attention-based model for high-resolution microscopy image analysis. Analogous to how pathologists examine slides under the microscope, our model utlizes weighted features from the entire slide for its classification. We showed that our model marginally outperforms the current sliding window method on a dataset of esophagus tissue with four classes of normal, BE-no-dysplasia, BE-with-dysplasia, and adenocarcinoma. Previous methodology for analyzing microscopy images is limited by bounding box annotations and unscalable heuristics. Our model, on the other hand, is trained end-to-end with only labels at the tissue-level, removing the high cost of data annotation and creating new opportunities for the use of deep learning in digital pathology.

# ACKNOWLEDGMENTS

The authors would like to thank Lamar Moss and Maksim Bolonkin for their help with this manuscript.

# FUNDING

This research was supported in part by grants R01LM012837 and P20GM104416 from the National Institutes of Health.

# REFERENCES

1. Haggitt RC. Barrett's esophagus, dysplasia, and adenocarcinoma. *Human pathology.* 1994;25(10):982-993.
2. Wild CP, Hardie LJ. Reflux, Barrett's oesophagus and adenocarcinoma: burning questions. *Nature Reviews Cancer.* 2003;3(9):676.
3. Conio M, Filiberti R, Blanchi S, et al. Risk factors for Barrett's esophagus: a case-control study. *International journal of cancer.* 2002;97(2):225-229.
4. Stein H, Siewert J. Barrett's esophagus: pathogenesis, epidemiology, functional abnormalities, malignant degeneration, and surgical management. *Dysphagia.* 1993;8(3):276-288.
5. Polednak AP. Trends in survival for both histologic types of esophageal cancer in US surveillance, epidemiology and end results areas. *International journal of cancer.* 2003;105(1):98-100.
6. Bollschweiler E, Wolfgarten E, Gutschow C, Hölscher AH. Demographic variations in the rising incidence of esophageal adenocarcinoma in white males. *Cancer: Interdisciplinary International Journal of the American Cancer Society.* 2001;92(3):549-555.
7. Blot WJ. Esophageal cancer trends and risk factors. *Seminars in oncology.* 1994;21(4):403-410.
8. Daly JM, Karnell LH, Menck HR. National Cancer Data Base report on esophageal carcinoma. *Cancer: Interdisciplinary International Journal of the American Cancer Society.* 1996;78(8):1820-1828.
9. Brown LM, Devesa SS. Epidemiologic trends in esophageal and gastric cancer in the United States. *Surgical oncology clinics of North America.* 2002;11(2):235-256.
10. Edgren G, Adami H-O, Weiderpass E, Nyrén O. A global assessment of the oesophageal adenocarcinoma epidemic. *Gut.* 2013;62(10):1406-1414.
11. Paull A, Trier JS, Dalton MD, Camp RC, Loeb P, Goyal RK. The histologic spectrum of Barrett's esophagus. *New England Journal of Medicine.* 1976;295(9):476-480.
12. Coco DP, Goldblum JR, Hornick JL, et al. Interobserver variability in the diagnosis of crypt dysplasia in Barrett esophagus. *The American journal of surgical pathology.* 2011;35(1):45-54.
13. Korbar B, Olofson AM, Miraflor AP, et al. Looking Under the Hood: Deep Neural Network Visualization to Interpret Whole-Slide Image Analysis Outcomes for Colorectal Polyps. *Computer Vision and Pattern Recognition Workshops (CVPRW), 2017 IEEE Conference on.* 2017:821-827.
14. Hou L, Samaras D, Kurc TM, Gao Y, Davis JE, Saltz JH. Patch-based convolutional neural network for whole slide tissue image classification. *Proceedings of the IEEE Conference on Computer Vision and Pattern Recognition.* 2016:2424-2433.
15. Cosatto E, Laquerre P-F, Malon C, et al. Automated gastric cancer diagnosis on h&e-stained sections; ltraining a classifier on a large scale with multiple instance machine learning. *Medical Imaging 2013: Digital Pathology.* 2013;8676:867605.
16. Saha M, Chakraborty C, Racoceanu DJCMI, Graphics. Efficient deep learning model for mitosis detection using breast histopathology images. 2018;64:29-40.

17. Komura D, Ishikawa SJC, Journal SB. Machine learning methods for histopathological image analysis. 2018;16:34-42.
18. Xu J, Luo X, Wang G, Gilmore H, Madabhushi AJN. A deep convolutional neural network for segmenting and classifying epithelial and stromal regions in histopathological images. 2016;191:214-223.
19. Coudray N, Ocampo PS, Sakellaropoulos T, et al. Classification and mutation prediction from non–small cell lung cancer histopathology images using deep learning. 2018;24(10):1559.
20. Liu Y, Gadepalli K, Norouzi M, et al. Detecting cancer metastases on gigapixel pathology images. *arXiv preprint arXiv:170302442.* 2017.
21. Wei J, Wei J, Jackson C, Ren B, Suriawinata A, Hassanpour S. Automated detection of celiac disease on duodenal biopsy slides: A deep learning approach. *Journal of Pathology Informatics.* 2019;10(1):7-7.
22. Wei JW, Tafe LJ, Linnik YA, Vaickus LJ, Tomita N, Hassanpour S. Pathologist-level classification of histologic patterns on resected lung adenocarcinoma slides with deep neural networks. *Scientific Reports.* 2019;9(1):3358.
23. Korbar B, Olofson AM, Miraflor AP, et al. Deep learning for classification of colorectal polyps on whole-slide images. *Journal of pathology informatics.* 2017;8.
24. Xu K, Ba J, Kiros R, et al. Show, attend and tell: Neural image caption generation with visual attention. *International conference on machine learning.* 2015:2048-2057.
25. You Q, Jin H, Wang Z, Fang C, Luo J. Image captioning with semantic attention. *Proceedings of the IEEE conference on computer vision and pattern recognition.* 2016:4651-4659.
26. Chen L-C, Yang Y, Wang J, Xu W, Yuille AL. Attention to scale: Scale-aware semantic image segmentation. *Proceedings of the IEEE conference on computer vision and pattern recognition.* 2016:3640-3649.
27. Fu J, Zheng H, Mei T. Look closer to see better: Recurrent attention convolutional neural network for fine-grained image recognition. *CVPR.* 2017;2:3.
28. Wang F, Jiang M, Qian C, et al. Residual attention network for image classification. 2017.
29. Jaderberg M, Simonyan K, Zisserman A. Spatial transformer networks. *Advances in neural information processing systems.* 2015:2017-2025.
30. Guan Q, Huang Y, Zhong Z, Zheng Z, Zheng L, Yang YJapa. Diagnose like a radiologist: Attention guided convolutional neural network for thorax disease classification. 2018.
31. Pesce E, Ypsilantis P-P, Withey S, Bakewell R, Goh V, Montana GJapa. Learning to detect chest radiographs containing lung nodules using visual attention networks. 2017.
32. Ypsilantis P-P, Montana GJapa. Learning what to look in chest X-rays with a recurrent visual attention model. 2017.
33. Corredor G, Whitney J, Pedroza VLA, Madabhushi A, Castro ERJJoMI. Training a cell-level classifier for detecting basal-cell carcinoma by combining human visual attention maps with low-level handcrafted features. 2017;4(2):021105.

34. Long J, Shelhamer E, Darrell T. Fully convolutional networks for semantic segmentation. *Proceedings of the IEEE conference on computer vision and pattern recognition.* 2015:3431-3440.
35. Schlemper R, Riddell R, Kato Yea, et al. The Vienna classification of gastrointestinal epithelial neoplasia. 2000;47(2):251-255.
36. He K, Zhang X, Ren S, Sun J. Delving deep into rectifiers: Surpassing human-level performance on imagenet classification. *Proceedings of the IEEE international conference on computer vision.* 2015:1026-1034.
37. Srivastava N, Hinton G, Krizhevsky A, Sutskever I, Salakhutdinov RJTJoMLR. Dropout: a simple way to prevent neural networks from overfitting. 2014;15(1):1929-1958.
38. Ioffe S, Szegedy CJapa. Batch normalization: Accelerating deep network training by reducing internal covariate shift. 2015.
39. Glorot X, Bengio Y. Understanding the difficulty of training deep feedforward neural networks. *Proceedings of the thirteenth international conference on artificial intelligence and statistics.* 2010:249-256.
40. Krizhevsky A, Sutskever I, Hinton GE. Imagenet classification with deep convolutional neural networks. *Advances in neural information processing systems.* 2012:1097-1105.
41. Smith LN. Cyclical learning rates for training neural networks. *Applications of Computer Vision (WACV), 2017 IEEE Winter Conference on.* 2017:464-472.
42. Loshchilov I, Hutter FJapa. Sgdr: Stochastic gradient descent with warm restarts. 2016.
43. Paszke A, Gross S, Chintala S, Chanan G. PyTorch. In:2017.
44. Odze RD, Goldblum JR. *Odze and Goldblum Surgical Pathology of the GI Tract, Liver, Biliary Tract and Pancreas E-Book.* Elsevier Health Sciences; 2014.

**LEGENDS**

**Figure 1**. Overview of our attention-based model. (a) An input image is divided into r × c grid cells (dividing lines are shown only for visualization). (b) Features extracted from each grid cell build a grid-based feature map tensor U. (c) Learnable 3-dimensional convolutional filters of size k × d × d (where d denotes the height and width of the convolutional filters) are applied on U feature map to generate an attention map α, which operates as the weights for an affine combination of feature vectors in U. The α represents a 2-dimensional attention map whose size is r in height and c in width; CNN, convolutional neural network; r and c, the number of rows and columns of input tissue grid; U, a tensor of features extracted from each grid cell, and its size is r in height, c in width, and k in depth; and z, a vector of features representing a whole-input image.

**Figure 2**. The confusion matrix for different histologic classes related to esophageal cancer compares the classification agreement of our attention-based model with pathologists' consensus.

**Figure 3**. Receiver operating characteristic curves for (a) the sliding window approach and (b) our proposed attention-based method.

**Figure 4**. Examples of attention maps generated by an attention module that is optimized for attending to the features of the adenocarcinoma class. Top row shows whole-slide images from the test set. The second row shows attention maps of the selected attention module for input images from four ground truth classes: (a) normal, (b) BE-no-dysplasia, (c) BE-with-dysplasia, and (d) adenocarcinoma. Higher attention weight is denoted by white color and lower is denoted by black color. For visualization purposes, each map is normalized so its maximum value is 1. The accuracy of attended regions for the adenocarcinoma class images are verified qualitatively by two expert pathologists. In contrast, the attention module is inattentive to lower risk class images.

# Supplementary Materials

**eMethods 1.** Details of Our Data Annotation Procedure

**eMethods 2.** Details of Our Attention-Based Deep Learning Architecture

**eFigure 1.** Typical Examples of a Whole-Slide Image and Class-Associated Patches

**eFigure 2.** Additional Examples of Visualized Attention Maps Attending Adenocarcinoma Class Features

**eTable.** Class Distribution of Images in Our Dataset

**eMethods 1.** Details of Our Data Annotation Procedure

In this study, two expert pathologists from the Department of Pathology and Laboratory Medicine at DHMC annotated each whole-slide image by drawing the smallest rectangular bounding boxes around characteristic lesions of each class in each image using Aperio ImageScope software and its Rectangle Tool. The marked ROIs are then extracted as cropped images in JPEG format.

The bounding box annotation is suitable in this study because this method is able to capture the histology patterns without well-defined boundaries, which is suitable for the diagnosis of Barrett's Esophagus based on continuous pathologic patterns. In addition, it reduces the annotation cost on our pathologists. In terms of the costs vs. benefits, a polygon-based annotation is suitable for dense predictions (e.g., a segmentation task), while a bounding box annotation is less demanding and is widely used for classification tasks due to its convenience and robustness.

**eMethods 2.** Details of Our Attention-Based Deep Learning Architecture

**Grid-based Feature Extraction**

To extract features on a high-resolution image through a CNN, we first divide the input image into smaller tiles with no overlap (Figure 1a), and then apply a CNN-based feature extraction on each tile (i.e., grid cell) of an r×c grid, with a $k$ feature vector extracted from each cell, resulting in the formation of a structured grid-based feature map $U$ of size k×r×c (Figure 1b). This feature map $U$ is a high-level feature expression of a high-resolution image while preserving the geometric relationships of local features. While the grid-based approach is robust even if full view of a lesion is not in a grid cell due to training with tissue-level geometric augmentation (e.g., random rotation and translation), the granularity of analysis can be further controlled by using overlapping tiles at a higher computational cost. Whereas existing methodology makes a crop prediction solely based on a crop and later aggregates prediction results of crops to build a whole image prediction, our feature structure enables us to directly analyze the whole image through an attention mechanism, which we present in the next subsection below.

In the implementation of CNN architecture for feature extraction, we use the residual neural network (ResNet) architecture,[1] one of the state-of-the-art CNN models with high performance on the ImageNet Large Scale Visual Recognition Competition (ILSVRC) as well as many medical image classification tasks.[2-4] Among several variants of ResNet models, we choose the pre-activation ResNet-18 model.[5] This model achieves a good trade-off between performance and GPU memory usage, which is vital for processing high-resolution images. By removing the final fully-connected layer before the global pooling layer, the network produces a 512-feature vector (k=512) as output for a tile input.

**Attention-based Classification**

After feature extraction, attention modules are applied to the feature map with their weights determining the importance or value of each tile in diagnostic relevancy (Figure 1c). The importance of each tile is estimated based on features extracted from the tile and its neighboring tiles because the adjoining areas of ROIs can also present informative characteristics. We compute a set of values, $V \in R^{r \times c}$, for a grid. To implement this local valuation function in a deep learning framework while maintaining the robustness for an arbitrary size of grid input, we utilize 3D convolutional filters of size k×d×d, where $k$ corresponds to the size of features and $d$ denotes the height and width of the kernels. In this framework, applying a 3D convolution kernel to a feature map $U$ generates a grid of value estimation $V$. We normalize $V$ by applying a softmax function to build an attention map $\alpha$, where $i$ and $j$ are row and column indices:

$$\alpha(V)_{i,j} = \frac{e^{V_{i,j}}}{\sum_{h=1}^{r}\sum_{w=1}^{c} e^{V_{h,w}}} \quad (1)$$

This attention map shows the relative importance of each tile and thus we compute a whole-slide global feature vector using the attention map. Specifically, by treating the attention map $\alpha$ as feature weights, the $n$-th components of the final feature vector $z$ are computed as follows:

$$z_n = \sum_{h=1}^{r}\sum_{w=1}^{c} \sigma(V)_{h,w} \cdot U_{n,h,w} \quad (2)$$

The feature vector $z$ is subsequently used for whole-slide classification through fully connected layers and a non-linear activation function, allowing for classification of the entire whole-slide image by optimizing for a label.

Moreover, the use of multiple attention modules in our framework can potentially capture more local patterns for classification, increasing the capacity and robustness of the network, especially for medical images of high resolution. As such, we simultaneously apply $m$ 3D filters that generate $m$ attention maps and individually populate $m$ feature vectors. All feature vectors are concatenated to form a single vector, which is fed to the fully connected classifier.

# eReferences

1. He K, Zhang X, Ren S, Sun J. Deep residual learning for image recognition. *Proceedings of the IEEE conference on computer vision and pattern recognition.* 2016:770-778.
2. Khosravi P, Kazemi E, Imielinski M, Elemento O, Hajirasouliha IJE. Deep convolutional neural networks enable discrimination of heterogeneous digital pathology images. 2018;27:317-328.
3. Chung Y-A, Weng W-HJapa. Learning Deep Representations of Medical Images using Siamese CNNs with Application to Content-Based Image Retrieval. 2017.
4. Zhang Z, Xie Y, Xing F, McGough M, Yang L. Mdnet: A semantically and visually interpretable medical image diagnosis network. *Proceedings of the IEEE conference on computer vision and pattern recognition.* 2017:6428-6436.
5. He K, Zhang X, Ren S, Sun J. Identity mappings in deep residual networks. *European conference on computer vision.* 2016:630-645.

**eFigure 1.** Typical Examples of a Whole-Slide Image and Class-Associated Patches

(a) A typical whole-slide image in our dataset. This particular slide contains three separate tissues and is of size 9,440 × 15,340 pixels. (b) Samples from each histology class in our dataset.

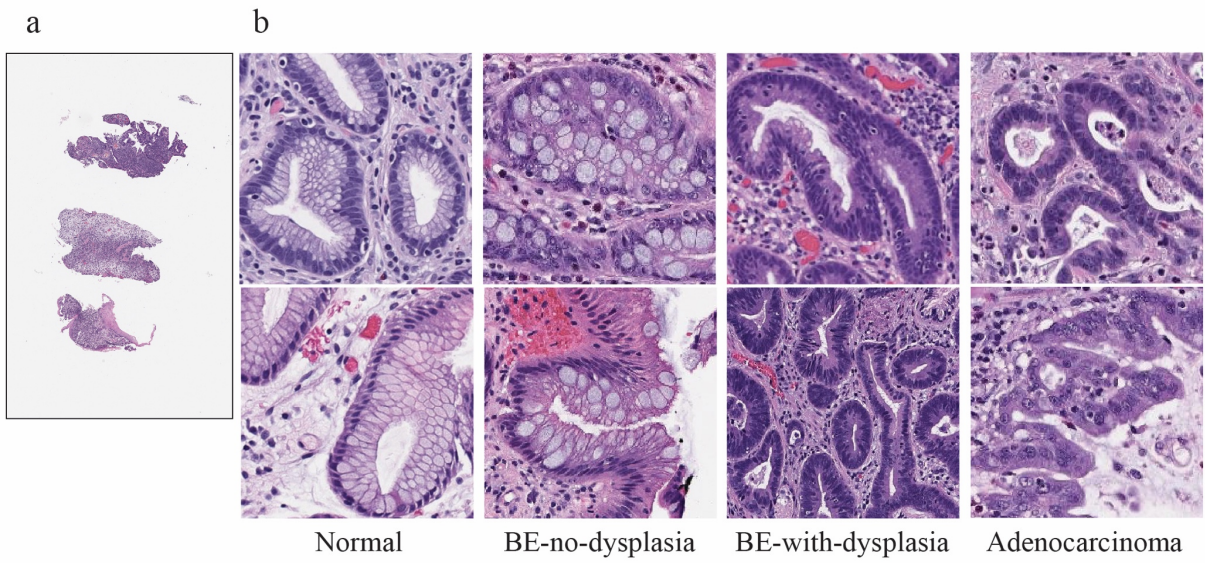

Normal　　BE-no-dysplasia　　BE-with-dysplasia　　Adenocarcinoma

**eFigure 2.** Additional Examples of Visualized Attention Maps Attending to Adenocarcinoma Class Features

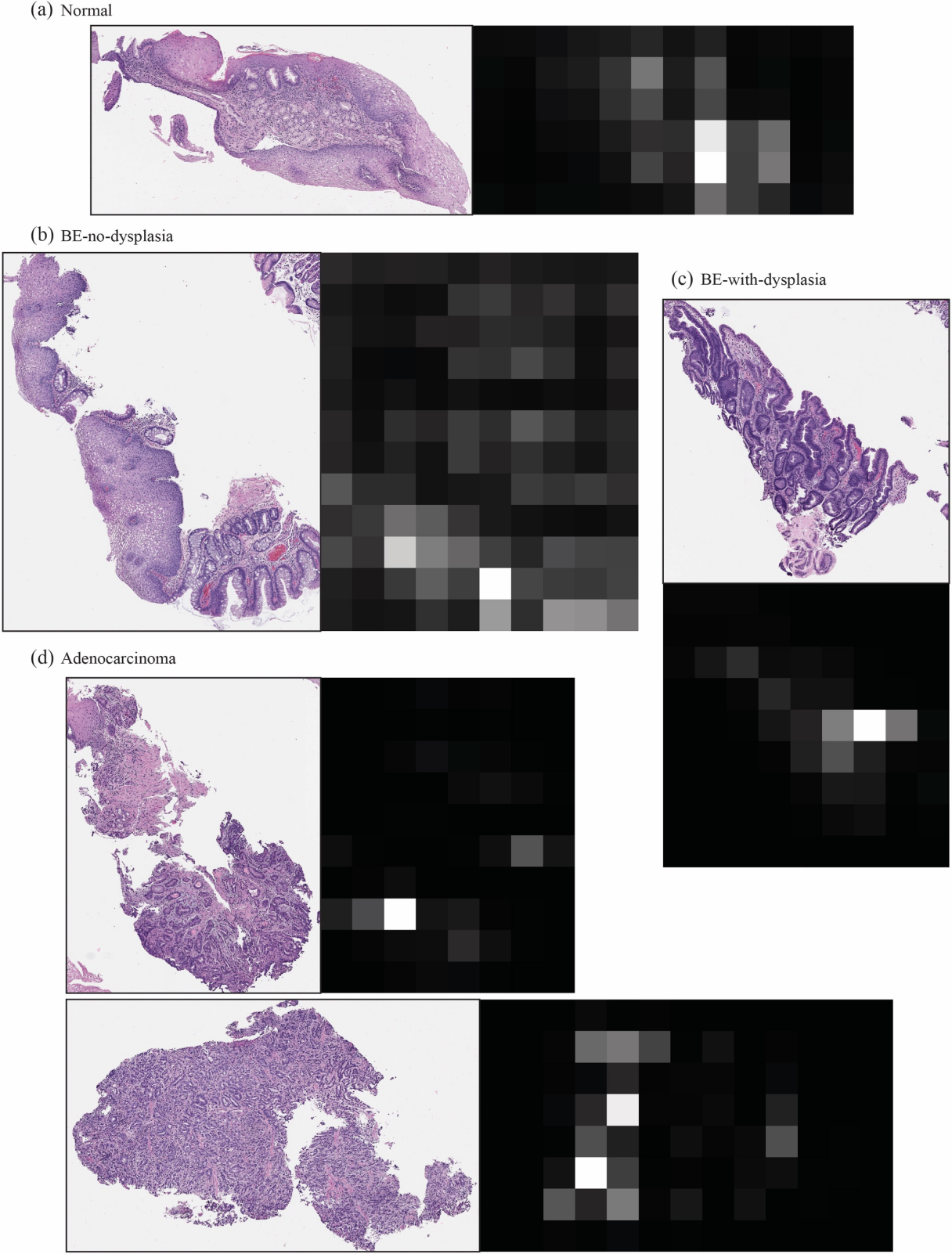

(a) Normal

(b) BE-no-dysplasia

(c) BE-with-dysplasia

(d) Adenocarcinoma

**eTable.** Class Distribution of Images in Our Dataset

| Diagnosis | Number (%) | | |
| --- | --- | --- | --- |
| | Training | Validation | Test |
| Normal | 115 (56.1%) | 22 (43.1%) | 58 (47.2%) |
| BE-no-dysplasia | 37 (18.0%) | 13 (25.5%) | 30 (24.4%) |
| BE-with-dysplasia | 23 (11.2%) | 9 (17.6%) | 14 (11.4%) |
| Adenocarcinoma | 30 (14.6%) | 7 (13.7%) | 21 (17.1%) |